\documentclass[conference]{IEEEtran}
\usepackage{algorithm,amsbsy,amsmath,amssymb,epsfig,bbm,mathrsfs,multirow,amsthm}
\usepackage{color}
\usepackage{graphicx,caption,subcaption,array,multirow,bm}%for three figures side-by-side
\usepackage{makecell}
\usepackage{cite}
\usepackage{threeparttable}
\usepackage{algorithmic}
\usepackage{xurl}
\DeclareMathAlphabet{\mathpzc}{OT1}{pzc}{m}{it}

\makeatletter
\def\endthebibliography{%
	\def\@noitemerr{\@latex@warning{Empty `thebibliography' environment}}%
	\endlist
}
\makeatother
\usepackage[hyphenbreaks]{breakurl}

\def\BibTeX{{\rm B\kern-.05em{\sc i\kern-.025em b}\kern-.08em
    T\kern-.1667em\lower.7ex\hbox{E}\kern-.125emX}}
\begin{document}

%\title{Metaverse Slicing: Multi-tiers Resource Architecture Management for Metaverse}
\title{Dynamic Resource Allocation for Metaverse Applications with Deep Reinforcement Learning}
\author{\IEEEauthorblockN{
		Nam~H.~Chu \textsuperscript{1},					
		Diep~N.~Nguyen\textsuperscript{1},
		Dinh~Thai~Hoang\textsuperscript{1},			
		Khoa~T.~Phan\textsuperscript{2},\\		
		Eryk~Dutkiewicz\textsuperscript{1}, 
		Dusit Niyato\textsuperscript{3},
		and Tao~Shu\textsuperscript{4}\\}
	\textsuperscript{1}School of Electrical and Data Engineering, University of Technology Sydney, Australia\\
	\textsuperscript{2}School of Engineering and Mathematical Sciences, La Trobe University, Melbourne, Australia\\
	\textsuperscript{3}School of Computer Science and Engineering, Nanyang Technological University, Singapore\\
	\textsuperscript{4}Department of Computer Science and Software Engineering, Auburn University, USA\\
}	

\maketitle

\begin{abstract} 
This work proposes a novel framework to dynamically and effectively manage and allocate different types of resources for Metaverse applications, which are forecasted to demand massive resources of various types that have never been seen before.
	Specifically, by studying functions of Metaverse applications, we first propose an effective solution to divide applications into groups, namely MetaInstances, where common functions can be shared among applications to enhance resource usage efficiency.
	Then, to capture the real-time, dynamic, and uncertain characteristics of request arrival and application departure processes, we develop a semi-Markov decision process-based framework and propose an intelligent algorithm that can gradually learn the optimal admission policy to maximize the revenue and resource usage efficiency for the Metaverse service provider and at the same time enhance the Quality-of-Service for Metaverse users.
	Extensive simulation results show that our proposed approach can achieve up to 120\% greater revenue for the Metaverse service providers and up to 178.9\% higher acceptance probability for Metaverse application requests than those of other baselines.
\end{abstract}


\section{Introduction}
\label{sec:introduction}
Although the Metaverse paradigm was first introduced in 1992, the recent efforts from big companies (e.g., Facebook, NVIDIA, and Microsoft) have made the Metaverse to be one of the most active fields for both academia and industry~\cite{xu2022full}. 
	The major difference between the Metaverse and existing virtual worlds (e.g., Pokemon Go and Second Life) is that Metaverse can be recognized as the seamless integration of multiple virtual worlds~\cite{xu2022full}. 
	Specifically, conventional virtual worlds limit the users' experiences in their specific environment, whereas the Metaverse allows users to seamlessly move between virtual worlds by a unified representation (e.g., an avatar).
	Therefore, similar to our real lives, users in Metaverse can preserve their belongings and appearance regardless of their application that they are experiencing. 
	Moreover, the Metaverse is likely to blend the physical world and the digital world with the aid of innovative technologies, such as Digital Twin and Extended Reality (XR)~\cite{xu2022full}.
	Thus, the Metaverse is expected to revolutionize various aspects of our lives, such as healthcare, smart industries, and e-commerce.

However, to fulfil the user experience and Quality-of-Service (QoS) requirements, the Metaverse indeed requires extremely intensive resource demands that have never been seen before.
	First, the comprehensive integration of XR in the Metaverse demands enormous data collected from perceived networks, e.g., the Internet of Things (IoT),  intensive computing for rendering three-dimensional objects, and ultra-high-speed and low-latency connections for guaranteeing the QoS and seamless user experience.	
	Second, millions of users are expected to join the Metaverse concurrently, making the network data usage increase more than $20$ times~\cite{credit_metaverse_2022}.
	Third, the Metaverse puts new constraints on networking. 
	In particular, in the current online system (e.g., massive multiplayer online games), the downlink requires a much higher throughput than that of the uplink~\cite{wang_characterizing_2012}.
	In contrast, the Metaverse likely demands intensively high throughput for both down and up connections.   
	It stems from the fact that users can create, share, and trade their assets to other users in any virtual world in the Metaverse. 
	Thus, the Metaverse will demand paramount resources exceeding those of any existing online platform~\cite{xu2022full}.
	As a result, resource management in Metaverse is one of the biggest challenges hindering the deployment of Metaverse. 

To address the Metaverse resource management, utilizing the multi-tier cloud computing architect can be considered as a promising solution.
First, multi-tier computing can relief the burden of massive resource demands on end-users for running Metaverse applications.
	Second, the multi-tier architecture can mitigate point-of-congestion problems of the centralized computing resource allocation architecture, where all resources are gathered and allocated from a centralized node.
	Third, the distribution of resources (e.g. computing, storage, and networking) along the path from end-users to the cloud can reduce the stress due to the enormous amount of data exchanged by the Metaverse operation over the networks.
	Finally, moving resources nearer to end-users results in decreasing delay, which is one of the most important factors in user experiences~\cite{zhao_estimating_2017}.
	Thus, the multi-tier computing architecture can be considered to be the most suitable solution for the Metaverse.

Since the Metaverse is only at the beginning stage, only a few research works consider resource management~\cite{jiang2021reliable, xu2021wireless, ng2021unified, han2021dynamic}.
	In~\cite{jiang2021reliable}, the authors address the problem of allocating resources by considering an edge-computing architecture to allocate computing resource to nearby Metaverse users.
	%	In~\cite{xu2021wireless} and \cite{ng2021unified}, the work in~\cite{jiang2021reliable} is extended by considering more types of resources (e.g., storage, radio).
	The work in~\cite{jiang2021reliable} is extended in ~\cite{xu2021wireless} and \cite{ng2021unified} by considering different types of resources (e.g., storage and radio). 
	Differently, the authors in~\cite{han2021dynamic} consider the resource allocation for perception networks (e.g., the IoT) that are employed to get real-world data for the Metaverse applications.
	Note that none of the above works investigates the multi-tier computing architecture for resource allocation in Metaverse.
	Instead, they only consider a single-tier computing architecture, which is unable to facilitate the Metaverse's massive resource demand.
	In addition, it can be observed that applications of Metaverse may share some same functions.
	For example, in practice, many applications, e.g., the Walking Dead: Our World and Pokemon Go, are currently using the same functions provided by the Google Maps API~\cite{pokemon}.    
	Thus, resource utilization can increase dramatically if a common function can be shared between applications.
	Nevertheless, all of the above studies are unable to take advantage of this aspect to maximize resource usage efficiency.
	Moreover, the resource demands of Metaverse are highly dynamic and uncertain since users can come and leave at any time, making  resource management based on conventional optimization methods ineffective.
	Therefore, effective solutions to address these problems are urgently needed. 
	
To address the above challenges, this paper proposes a novel framework that can automatically and intelligently manage various resource types of the underlying multi-tier computing architecture to maximize the performance of the Metaverse system.
	First, we propose a new application decomposition technique for Metaverse applications, by which functions of a Metaverse application can be executed separately at different tiers of the computing architecture depending on the available resources of each tier and the requirements of these functions.
	As such, this technique can leverage resources at different tiers simultaneously, thereby providing a flexible and high efficient solution for managing Metaverse applications.
	Second, we propose a novel paradigm, namely MetaInstance, that can exploit the similarities of Metaverse applications to maximize resource utilization. 
%	In particular, applications in the Metaverse system will be divided into multiple MetaInstances such that applications in one MetaInstance can share some common functions, thereby possibly saving more resources.
	Third, we develop a highly-effective framework based on the semi-Markov decision process to capture the real-time property of the Metaverse together with a self-learning algorithm based on deep reinforcement learning to automatically learn the optimal policy for the system under the resource demand's uncertainty and dynamic.
	Finally, we perform extensive simulation and show that our proposed solution clearly outperforms other baseline approaches. 
%	our solution can achieve up to 120\% greater revenue for the Metaverse service providers and up to 178.9\% higher acceptance probability for Metaverse application requests than those of other baseline methods.
%	to evaluate the proposed solution, we perform simulations, whose results show that our proposed solution clearly outperforms other baseline approaches by up to 69.9\% greater revenue.

\section{Dynamic Multi-tiers Resource Allocation Architecture for Metaverse}
\label{sec:model}
\begin{figure}[t]
	\centering
	\includegraphics[width=0.95\linewidth]{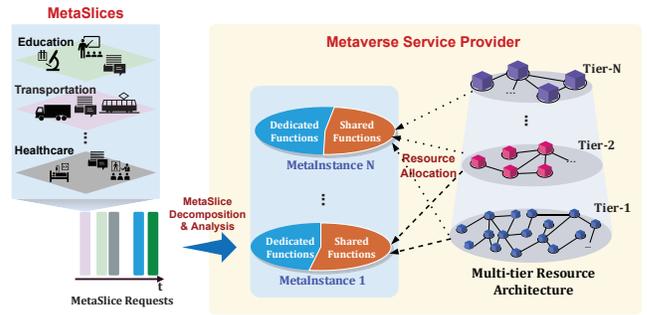}
	\caption{The system model of the proposed framework, in which various resource types at different tiers can be simultaneously utilized and shared to create MetaSlices.}
	\label{fig.system_model}
	\vspace{-15pt}
\end{figure}
As illustrated in Fig.~\ref{fig.system_model}, this work considers a Metaverse system managed by a service provider, where multiple Metaverse applications (e.g., healthcare and education) can operate simultaneously.
	To address the massive resource demands of Metaverse, we propose a novel multi-tier resource management framework that can dynamically and effectively distribute various resource types for Metaverse's applications.
	Specifically, we consider that the system has $P$ types of resources (e.g., networking, storage, and computing), which are distributed at different tiers along the way from end-users to the cloud.
	By doing so, this framework can offer a more flexible, efficient, and robust solution for deploying Metaverse applications compared to that of the centralized cloud architecture.
It can be observed that a Metaverse application, namely MetaSlice, consists of independence functions, and thus we propose an application decomposition technique, in which functions of an application can be separately created at different tiers depending on their requirement and the available resources at each tier.
	For example, a MetaSlice for navigation may include a driving assistant, real-time traffic, and a digital map.
	Since the driving assistant and real-time traffic require low delay, they can be created at a low tier (e.g., tier-1).
	In contrast, the digital map has a low update frequency, and thus it can be created at a higher tier.
	Thus, the application decomposition can provide an effectively and flexibly way for deploying MetaSlices.
%	To enable this technique, we consider that a MetaSlice's deployment is instructed by the MetaBlueprint, i.e., a template defining the structure, configuration, and workflow for creating and managing this MetaSlice. 
	The decomposition of applications can be done by existing methods, e.g.,~\cite{alturki2019exploring}.
%	From the technical aspect, a MetaSlice's implementation can be considered similar to the network slice concept in the fifth-generation mobile network (5G)~\cite{ngmn_nsi_2016}. 

We consider that there are $G$ classes of MetaSlices according to their characteristics, e.g., QoS, user experience, and technical requirements.
	In addition, the system can host various MetaSlice types simultaneously, e.g., education, navigation, and tourism.
	As analyzed in the previous section, concurrent MetaSlices in the system may share some common functions.
	If common functions are shared among MetaSlices, a lot of resources can be saved, thereby increasing system resource usage efficiency and revenue of the provider.
	To that end, this study proposes to group MetaSlices into clusters, i.e., MetaInstances.
	As illustrated in Fig.~\ref{fig.system_model}, a MetaInstance can be determined by shared functions and dedicated functions of specific MetaSlices.
	To manage MetaSlices, we consider that each MetaInstance maintains a function description, which is updated whenever a MetaSlice is added or departs.
	Technically, a MetaSlice can be deployed in a similar way as as that of a network slice in 5G network slicing~\cite{ngmn_nsi_2016}.
	However, their functions and the ways they are implemented are very different. 
	Specifically, the 5G network slicing aims to provide various types of end-to-end connections from mobile users to service providers, e.g., ultra-reliable low-latency communications (uRLLC) and enhanced Mobile Broadband (eMBB).
	In network slicing, a network slice is a logical network built on top of a physical network to support one connection type (e.g., uRLLC and eMBB).
	Each network slice is created based on multiple predefined functions of 5G network providers.
	On the other hand, our proposed solution offers an effective and flexible approach to implementing and managing the Metaverse applications.
	In particular, our proposed framework first decomposes a MetaSlice (i.e., Metaverse application) into independent functions.
	Then, each function will be allocated different types of resources (e.g., radio, computing, and storage) at different network tiers regarding its requirements.
	In addition, because functions' serving capabilities are limited, the user experience may be degraded if too many MetaSlices share the same function.
	Therefore, this work considers that the maximum of $N_L$ MetaSlices can share one function simultaneously.
The above analysis demonstrates that the proposed framework can not only benefit the provider by maximizing resource usage efficiency but also offer a better user experience and QoS for end-users.
	To obtain these results, MetaSlice admission control and resource management play the key roles.
	On the one hand, accepting/rejecting a MetaSlice request determines the provider's revenue and the user's QoS (e.g., service availability).
	On the other hand, better resource utilization can help the provider to save more resources required to host future MetaSlices, thereby increasing its long-term revenue.
	Thus, this paper focuses on the admission control and resource management in the proposed framework.

%\subsection{Admission and Resource Managements}
Upon a MetaSlice request arrives, the Metaverse system makes a decision (i.e., accept or reject) based on the system's available resources and the requested MetaSlice's required resources and class.
	If the request of MetaSlice $m$ is accepted, the system will examine the accepted MetaSlice to determine its similarities with ongoing MetaInstances as follows.
	Let the functions in a MetaSlice $m$ be denoted by a function vector \mbox{$\mathbf{f}_m\!\triangleq\!\{ F_m^f\}_{f=1}^K$} where $K$ is the total number of functions supported by the system and $F_m^f$ represents an appearance of function $f$ in this MetaSlice, i.e., \mbox{$F_m^f=1$} if MetaSlice $m$ uses function $f$, and \mbox{$F_m^f=0$}, otherwise. 
	Similarly, the function vector of MetaInstance $i$	is given as \mbox{$\mathbf{f}_i \!\triangleq\!\{ F_i^f\}_{f=1}^K$}.
	Then, the similarity index between MetaSlice $m$ and MetaInstance $j$ is calculated by any similarity function such as cosine and Jaccard~\cite{jaccard1912the}.
	Here, we use the Jaccard similarity function, denoted by $d_{\text{Jaccard}}$, which is defined as follows:
	\begin{equation}
		d_\text{Jaccard}(\mathbf{f}_m,\mathbf{f}_j) = \frac{\mathbf{f}_m\cdot\mathbf{f}_j}{||\mathbf{f}_m||^2 +||\mathbf{f}_j||^2 - \mathbf{f}_m\cdot\mathbf{f}_j},
	\end{equation}
	where $||\cdot||$ is the 2-norm of a vector and the nominator is the dot product of two vectors. 
	
After obtaining all the similarity indexes, the system will add MetaSlice $m$ to the MetaInstance that has the highest similarity index.
	If MetaSlice $m$ has dedicated functions, the system will allocate resources to create these functions.
	In a case that MetaSlice $m$ does not have any common function with ongoing MetaInstances, the system will create a new MetaInstance for it.
	Once a MetaSlice completes/departs, its occupied resources are released.
	In practice, the request arrival and MetaSlice departure processes are highly dynamic and uncertain.
	To address these challenges, we develop a semi-Markov decision process-based framework in the next section.

\section{Metaverse Admission Control Formulation}	
\label{sec:formulation}
This paper develops a semi-Markov Decision Process (sMDP) to enable the system to adaptively decide to accept/reject a request based on (i) its currently available resources, (ii) the requested MetaSlice's required resources, and (iii) its class, under the high dynamic and uncertainty of request arrival and MetaSlice departure processes.
	In addition, the sMDP makes decisions in a real-time manner, and thus it can capture the real-time admission control.
	The rest of this section will describe our proposed sMDP-based framework.

\subsection{State Space and Action Space}
Since the resources of the provider are limited, to maximize the provider revenue, it is necessary to consider the system's available resources and the required resources of requests.
	In addition, the class identification (i.e., class ID) of a requested MetaSlice is also important because each class can bring different income for the provider.
	Generally, the sMDP specifies decision epochs as time points where decisions are taken~\cite{tijms_a_2003}.
	As such, in this work, we can define the decision epochs as inter-arrival time between two consecutive MetaSlice requests.
	Therefore, the system state $s$ at a decision epoch can be defined as 	\mbox{$s\triangleq\left(\mathbf{n}_u, \mathbf{n}_m,g\right)$}, where $g$ is the class ID of the requested MetaSlice, and \mbox{$\mathbf{n}_u\!=\!\{n_u^p\}_{d=1}^P$} and \mbox{$\mathbf{n}_m\!=\!\{n_m^p\}_{d=1}^P$} are two vectors representing the system available resources and the required resources of a requested MeraSlice, respectively.
	Each coordinate of these vectors, i.e., $n_u^p$ and $n_m^p$, specifies the number of resources types $p$. 
	Thus, the system state space is given as follows:
	\begin{equation}
		\label{state_space}
		\begin{aligned}
			\mathcal{S} \triangleq & \Big\{(\mathbf{n}_u, \mathbf{n}_m, g): g \in \{1,\dots,G\};\\
			&n_u^d \text{ and } n_m^p \in \{0,\dots,N^p\} \forall p \in \{1,\dots, P\} \Big\},
		\end{aligned}	
	\end{equation}
	where $N^p$ is the total number of resources type $p$ of the provider.

Note that in our proposed sMDP, a state transition from state $s$ to state $s'$ only happens when an event arises, e.g., the arrival of a MetaSlice request.
	Let \mbox{$\mathbf{e}\!=\!\{e_g\}_{i=1}^G$}	denote the system event, where \mbox{$e_g\in \{-1,0,1\}$} indicates that (i) if \mbox{$e_g=1$}, a request class-$g$ arrives, (ii) if \mbox{$e_g=-1$} a MetaSlice class-$g$ departs, and (iii) \mbox{$e_g=0$}, otherwise.
	Consequently, we can derive the set of all possible events as follows:
	\begin{equation}
		\mathcal{E} \triangleq \big\{\mathbf{e}:e_g\in \{-1,0,1\}; \sum_{g=1}^{G}|e_g| \leq 1  \big\}.
	\end{equation}	
	For the case when there is no MetaSlice of any class departing or arriving, we can define a trivial event, i.e., \mbox{$\mathbf{e^*} \triangleq (0,\dots, 0)$}.
	
	Suppose that at state $s$ a request arrives, then the system must take an action $a_s$, which is to accept (i.e., \mbox{$a_s\!=\!1$}) or reject (i.e., \mbox{$a_s\!=\!0$}) this request to maximize the provider's long-term revenue.
	Therefore, we can define the action space at state $s$ as $\mathcal{A}_{s} \triangleq \{0,1\}$.

\subsection{Transition Probability and Immediate Reward Function}
To obtain the transition probabilities that the system transits from one state to another, we adopt the uniformization technique~\cite{tijms_a_2003}.
In practice, end-users join and leave the system at any time, which is unknown in advance.
Thus, this paper considers that the arrival process of requests class-$g$ and the departure process of MetaSlices class-$g$ follow the Poison distribution with mean $\lambda_g$ and the exponential distribution with mean $1/\mu_g$, respectively~\cite{tijms_a_2003}.
Let $x_g$ denote the number of ongoing MetaSlices class-$g$, we then can represent the number of all running MetaSlices by a vector \mbox{$\mathbf{x} \triangleq \{x_g\}_{g=1}^G$}. 
Given the above, parameters of uniformization are defined as:
\begin{align}
	z &= \max_{\mathbf{x}\in\mathcal{X}} \sum_{g=1}^{G}(\lambda_g + x_g\mu_g),\\
	z_{\mathbf{x}} &=  \sum_{g=1}^{G}(\lambda_g + x_g\mu_g),
\end{align}	
where $\mathcal{X}$ denotes the set of all possible values for $\mathbf{x}$.
Now, the occurrence probability of the next event $\mathbf{e}$ is given as follows. 
The probability of an arrival of request class-$g$ appears in $\mathbf{e}$ is $\lambda_g/z$.
The probability of a departure of MetaSlice class-$g$ appears in $\mathbf{e}$ is $x_g\mu_g/z$, and the probability that the next event is a trivial event \mbox{$\mathbf{e}^*$ is $1\!-\!z_{\mathbf{x}}/z$}.
Then, we can obtain the transition probabilities for the sMDP.

To maximize the long-term revenue for the provider, the immediate reward function needs to consider the revenue from leasing resources.
In addition, the proposed application decomposition and MetaInstance techniques can offer a higher resource utilization by sharing resources among the MetaSlices.
As such, accepting a MetaSlice with lower occupied resources will benefit the provider in the long run, and thus the resource occupation of a requested MetaSlice is another important factor. 
Thus, we can define the reward function as:
\begin{align}
	r(\mathbf{s},a_s) = \left\{
	\begin{array}{ll}
		r_g -  \sum_{p=1}^{P}w_p n^p_o, &\mbox{if $e_g\!=\!1$ and $a_s\! =\! 1$},\\
		0, &\mbox{otherwise},					
	\end{array}	\right.
	\label{eq:reward_function}
\end{align}
where $r_g$ is an income obtained by releasing resources for a MetaSlice class-$g$, and  $n^p_o$ is an amount of resources type $p$ occupied by this MetaSlice.
Here, the weights, i.e., ${w_p}_{p=1}^P$, define the tradeoff between these factors, which can be set based on the provider's business strategies. 
In~\eqref{eq:reward_function}, it can be observed that even MetaSlices have the same income (i.e., $r_g$), accepting the one with lower occupied resources gets a greater reward, thereby maximizing the provider's long-term revenue.

The objective of this work is to find an optimal admission policy $\pi^*$ for the system to maximize the long-term average reward function, i.e., $\mathcal{R}(\pi)$, as follows:
\begin{eqnarray} 
	\label{eq:average_reward}
	\max_\pi \quad	{\mathcal{R}}(\pi)	=	\lim_{T \rightarrow \infty} \frac{1}{T} \sum_{t=1}^{T} {\mathbb{E}} \left[ r_t (s_t, \pi(s_t)) \right],	
\end{eqnarray}
where $\pi(s_t)$ is the action that is selected at state $s_t$ at time $t$ according to the admission policy $\pi$ and $r_t$ is an immediate reward derived from~\eqref{eq:reward_function}.
In the next section, we discuss our proposed intelligent algorithm that can automatically and effectively find the optimal policy $\pi^*$ for the system.

\section{Intelligent MetaSlice Admission Control}
\label{sec:solution}
In practice, the request arrival and MetaSlice departure processes are unknown in advance because end-users can come and leave at any time.
	Therefore, it is ineffective to apply conventional optimization techniques to find an optimal admission policy for the system.
	In this context, Deep Q-learning techniques can help the system gradually learn an optimal admission policy without requiring complete information about the request arrival and MetaSlice departure processes. 
	However, conventional deep Q-learning techniques face the overestimation problem when estimating the optimal Q-values~\cite{hasselt_doubledeep_2016}, i.e., \mbox{$\mathcal{Q}^*(s, a)$}, %that is the value of taking action $a$ at state $s$, 
	thereby possibly leading to an unstable learning process or even resulting in a sub-optimal policy.
	To address this issue, we develop a highly effective Deep Reinforcement Learning (DRL)-based algorithm for the system, namely iMSAC, that adopts recent advanced techniques, i.e., the buffer replay mechanism, the dueling neural network architecture~\cite{wang_dueling_2016}, and the double Q-learning~\cite{hasselt_doubledeep_2016}.
	The iMSAC is described in details in Algorithm~\ref{alg:imsac}. %In the following, its main components are presented.
\begin{algorithm}[h]
	\caption{The iMSAC}
	\label{alg:imsac}
	\begin{algorithmic}
		\STATE Initialize $\epsilon$, buffer $\mathbf{M}$, and Q-network $\mathcal{Q}$ with random parameters $\theta$. 
		\STATE Create target Q-network $\bar{\mathcal{Q}}$ by cloning the Q-network.
		\FOR{\textit{step = 1 to T}}
		\STATE Get action $a_t$ following the $\epsilon$-greedy policy as follows:
		\begin{align}
			a_t\! =\! \left\{
			\begin{array}{ll}						
				\!\underset{a \in \mathcal{A}}{\text{argmax}}~\mathcal{Q} \mbox{($\mathbf{s}_t,a; \theta_t)$},&\mbox{with probability $1\!-\!\epsilon$},\\
				\!\mbox{random action $a\!\in\!\mathcal{A}$}, &\mbox{otherwise}.												
			\end{array}	\right.
			\label{eq:epsilon_greedy}
		\end{align}
		\STATE Execute $a_t$, then observe reward $r_t$ and next state $\mathbf{s}_{t+1}$.
		\STATE Store experience \mbox{$(\mathbf{s}_t, a_t, r_t, \mathbf{s}_{t+1})$} in  $\mathbf{M}$.
		\STATE Sample $\mathbf{M}$ randomly to get a mini-batch of experiences.	
		\STATE Using ~\eqref{eq:recontruct_Qfunction_mean} and~\eqref{eq:targetDQN} to get the Q-value and the target Q-value, respectively.
		\STATE Update $\theta$ based on SGD algorithm. 
		\STATE Decrease the value of $\epsilon$.
		\STATE Set $\bar{\theta}= \theta$ at every $C$ steps.
		\ENDFOR
	\end{algorithmic}
\end{algorithm}

In Algorithm~\ref{alg:imsac}, at time $t$, the MetaSlicing system is at state $s_t$ and performs an action $a_t$ derived from the $\epsilon$-policy, as in~\eqref{eq:epsilon_greedy}.
	Then, the system moves to a new state \mbox{$s_{t+1}$} and receives an immediate reward $r_t$.
	Since experiences, i.e., tuples \mbox{$<s_t,a_t,s_{t+1},r_t>$}, obtained in sMDP are highly correlated, using them directly to train the Deep Neural Network (DNN) may lead to a slow convergence speed~\cite{mnih_human_2015}.
	To mitigate this problem, we adopt the replay buffer mechanism, in which experiences are stored in a buffer $\mathbf{M}$. 
	Then, to train the DNN, a mini-batch of experiences is sampled randomly from $\mathbf{M}$. 
	In iMSAC, the DNN is leveraged for estimating the Q-values so that the input and output layers are set according to the state dimension (i.e., available resources, required resources and class ID of the incoming MetaSlice request) and the action dimension (i.e., accept and reject), respectively.
	By inputting a state $s$ in the DNN, estimated Q-values for all actions at state $s$ are obtained, each corresponding to a neuron in the output layer.
	To stabilize the learning process, we adopt the dueling architecture that divides the iMSAC's DNN into two streams, one to estimate the state-value function $\mathcal{V}(s)$ and another for estimating the advantage function $\mathcal{W}(s, a)$.
	It is worth noting that while $\mathcal{V}(s)$ determines how good to be at a state $s$, $\mathcal{W}(s, a)$ specifies the importance of action $a$ in comparison with others at state $s$.
	Let $\beta$ and $\zeta$ denote parameters of the $\mathcal{V}(s)$ and $\mathcal{W}(s,a)$ streams, respectively.
	Thus, the Q-value of performing action $a$ at state $s$ is estimated by the iMSAC's DNN as follows~\cite{wang_dueling_2016}:
	\begin{equation}
		\label{eq:recontruct_Qfunction_mean}
		\mathcal{Q}(s,a; \beta, \zeta) = \mathcal{V}({s}; \beta) + \Big(\mathcal{W}({s},a; \zeta) - \frac{1}{|\mathcal{A}_{s}|} \sum_{a'\in\mathcal{A}_{s}}\mathcal{W}({s},a';\zeta)\Big).		
	\end{equation}	

To mitigate the overestimation problem, we adopt the double Q-learning method that uses two identical DNNs, which are Q-network $\mathcal{Q}$ for action selections, and target Q-network $\bar{\mathcal{Q}}$  for evaluating action.
	Then, the target Q-value at time $t$ is given as follows:
	\begin{equation}
		\label{eq:targetDQN}
		Z_t = r_t(s_t,a_t) + \alpha \bar{\mathcal{Q}}\big(\mathbf{s}_{t+1}, \underset{a}{\operatorname{argmax}}\mathcal{Q}(\mathbf{s}_{t+1},a; \theta_t);\bar{\theta}_t\big),
	\end{equation}
	where $\theta$ and $\bar{\theta}$ denote the $\mathcal{Q}$'s and $\bar{\mathcal{Q}}$'s parameters, respectively. The importance of future rewards is reflected by the discount factor $\alpha$.
	Since the objective of training $\mathcal{Q}$ is minimizing the distance between the estimated Q-value and the target Q-value, we can define a loss function at time $t$ as follows:
	\begin{equation}
		\begin{aligned}
			\label{eq:lossfunction}
			\mathcal{L}_t(\theta_t) = \mathbb{E}_{({s},a,r,{s}')}\left[ \left( H_t
			-\mathcal{Q}({s},a;\theta_t)\right)^2\right],
		\end{aligned}
	\end{equation}
	where $\mathbb{E}[.]$ is the expectation regard with data points, i.e., \mbox{$({s},a,r,{s}')$}, in $\mathbf{M}$.
	In this work, we employ Stochastic Gradient Descent (SGD) to minimize the loss function $\mathcal{L}_t(\theta_t)$ because this method has lower computing complexity compared with that of conventional Gradient Descent while the convergence is still guaranteed~\cite{robbins1951stochastic}. 
	Note that to stabilize the learning process, as in~\cite{mnih_human_2015}, the target Q-network $\bar{\mathcal{Q}}$ is not updated at every time step.
	Instead, at every $C$ steps, $\bar{\theta}$ is cloned from $\theta$. 
 
%%%%%%%%%%%%%%%%%%%%%%%%%%%%%%%%%%%%%%%%%%%%%%%%%%%%%%%%%%%%%%%%%%%

\section{Performance Evaluation}
\subsection{Simulation Setting}
\begin{figure}[t]
	\centering
	\includegraphics[width=0.6\linewidth]{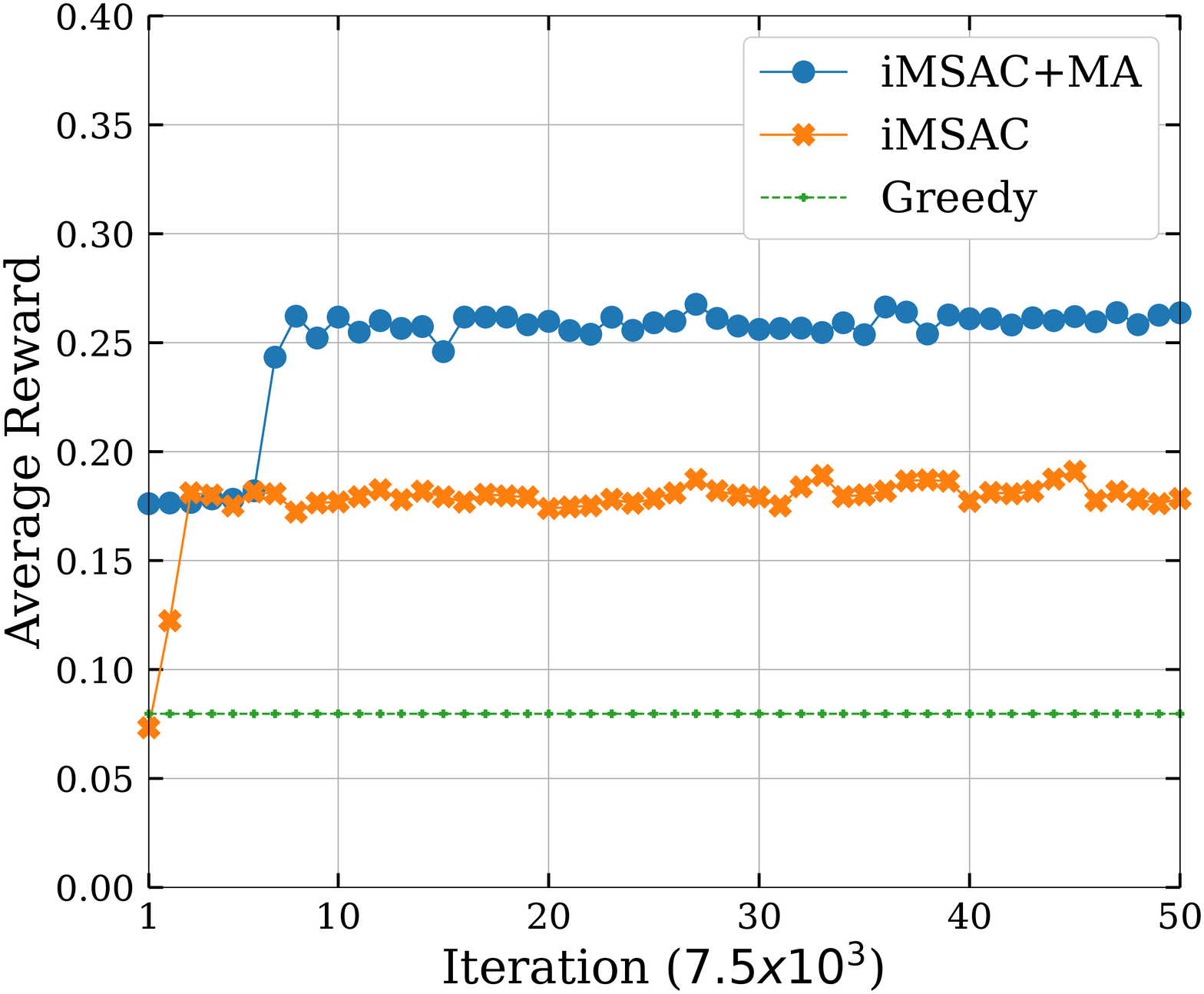} 
	\caption{Convergence rate of proposed algorithm iMSAC.}
	\label{fig:convergene}
	\vspace{-17pt}
\end{figure}

Unless otherwise stated, the simulation parameters are set as follows.
	The system supports nine function types.
	There are three different functions for each MetaSlice.
	We consider that a function can be shared among a maximum of five MetaSlices.
	MetaSlices are classified into three classes, i.e., class-1, class-2, and class-3.		
	The immediate reward $r_g$ is set to $1, 2$, and $4$, and $\lambda_g$ is set to $60, 40$, and $25$ request/hours for class-1, class-2, and class-3, respectively. 
	In the MetaSlice departure processes of all classes, $\mu_g$ is set at two MetaSlices/hours.
	We consider three resource types, i.e., radio, storage, and computing, and each MetaSlice's function requires similar resources as functions in the Network Slice~\cite{ghina_an_2020}, e.g., 40 MHz, 40 GB, and 40 GFLOPS/s.
	Note that the proposed algorithm iMSAC does not require complete information about surrounding environment (e.g., arrival and departure rates and total system resources) in advance.
	Instead, it can adapt its policy accordingly to practical demands and requirements.
	
\begin{figure}[t]
	\centering
	$\begin{array}{cccc}
		\includegraphics[width=0.45\linewidth]{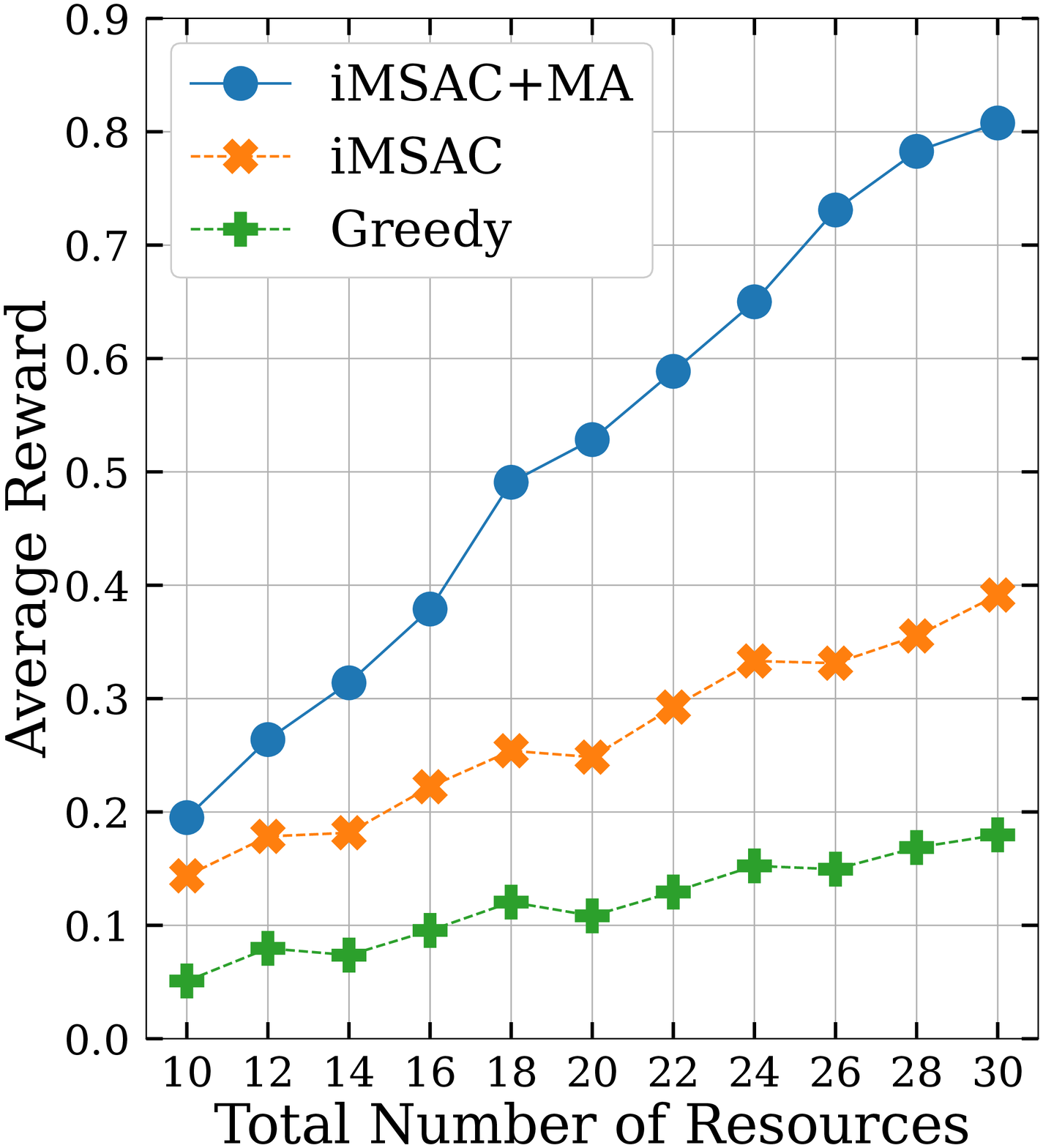}		
		&\includegraphics[width=0.45\linewidth]{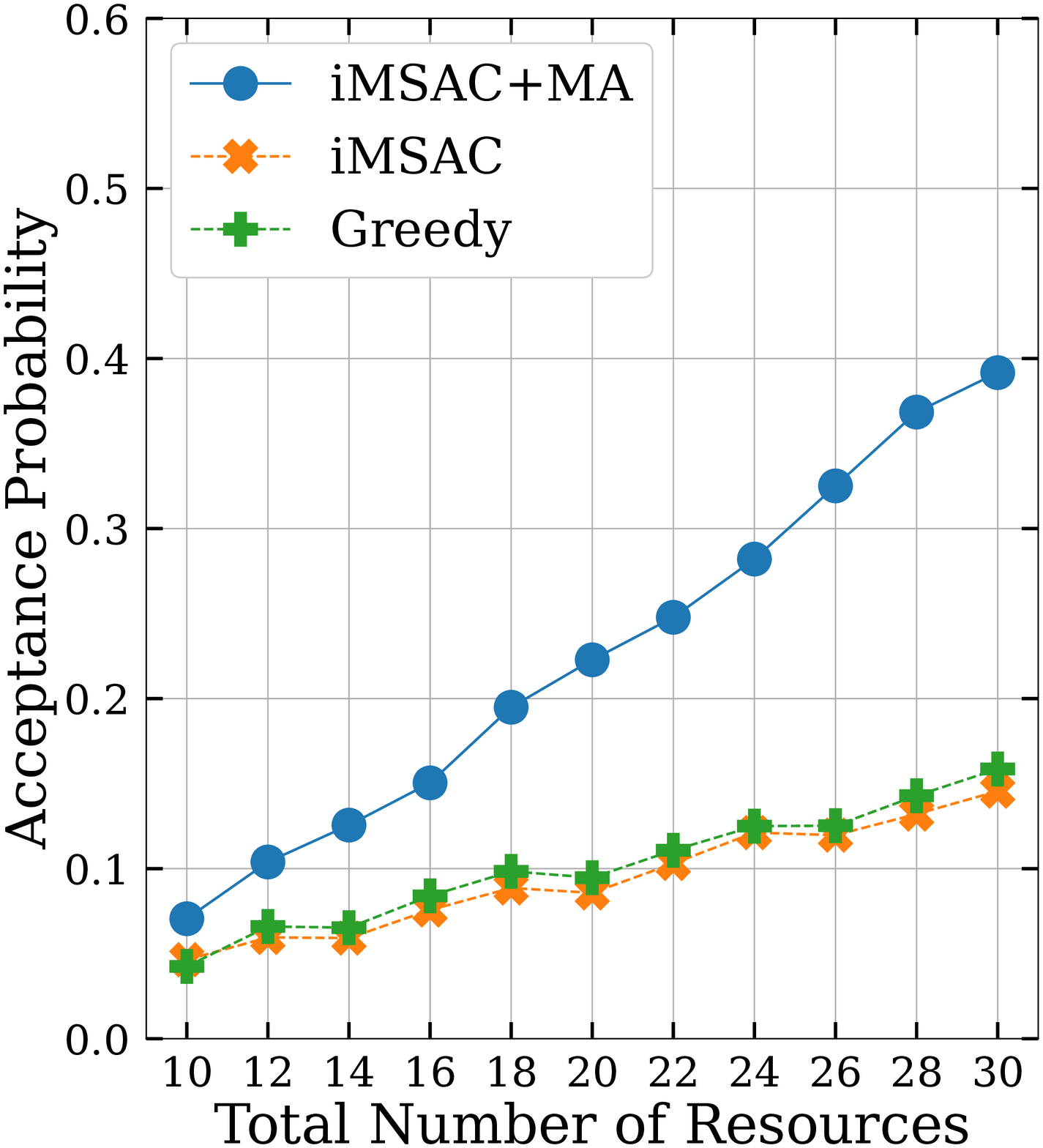}\\
		\text{(a) Average rewards} & \text{(b) Acceptance probability}
	\end{array}$
	\caption{Vary the total number of system resources.}
	\label{fig:vary_vms}
	\vspace{-16pt}
\end{figure}

\begin{figure*}[t]
	\centering
	$\begin{array}{cccc}
		\includegraphics[width=0.25\linewidth]{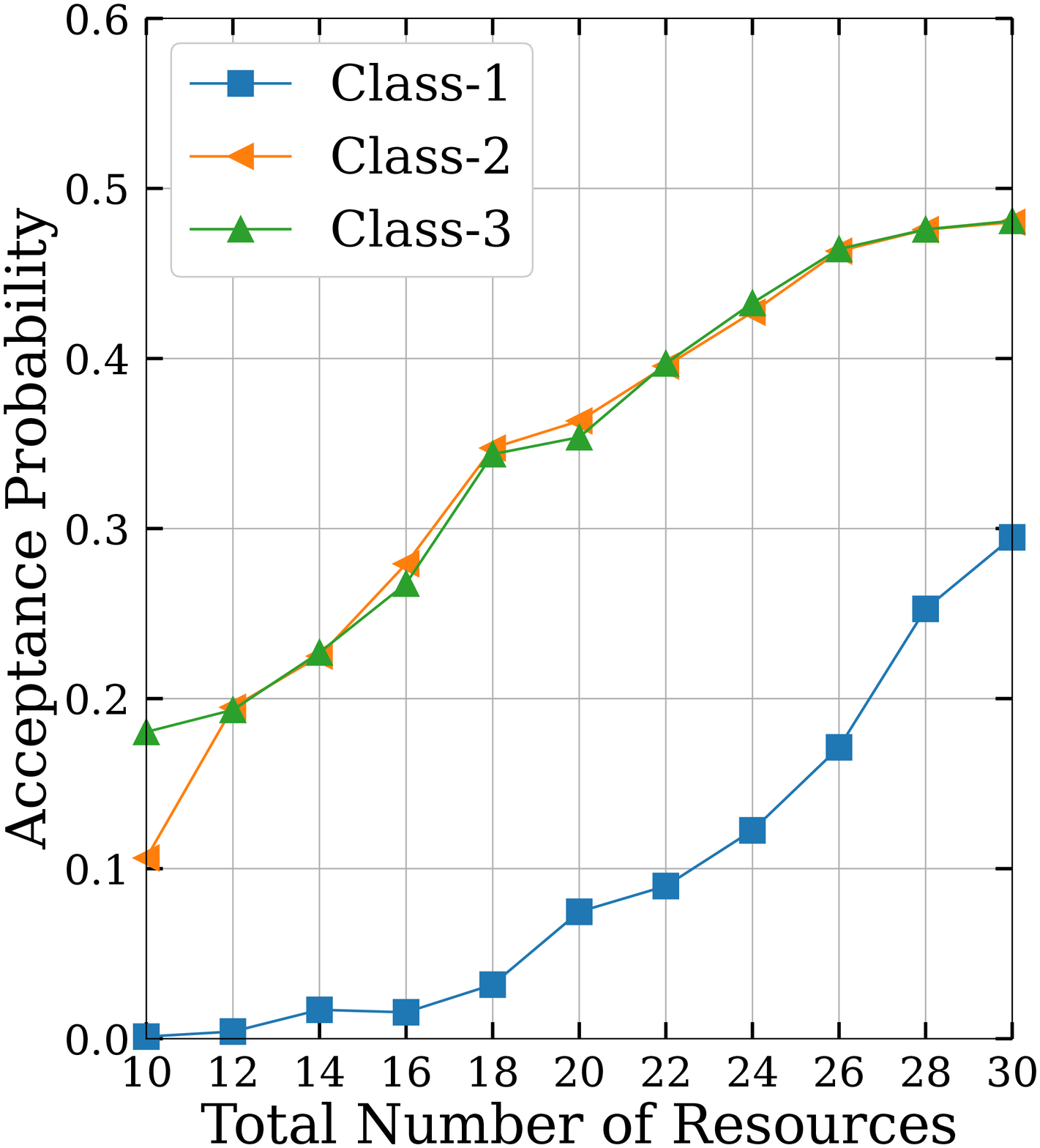}
		&\includegraphics[width=0.25\linewidth]{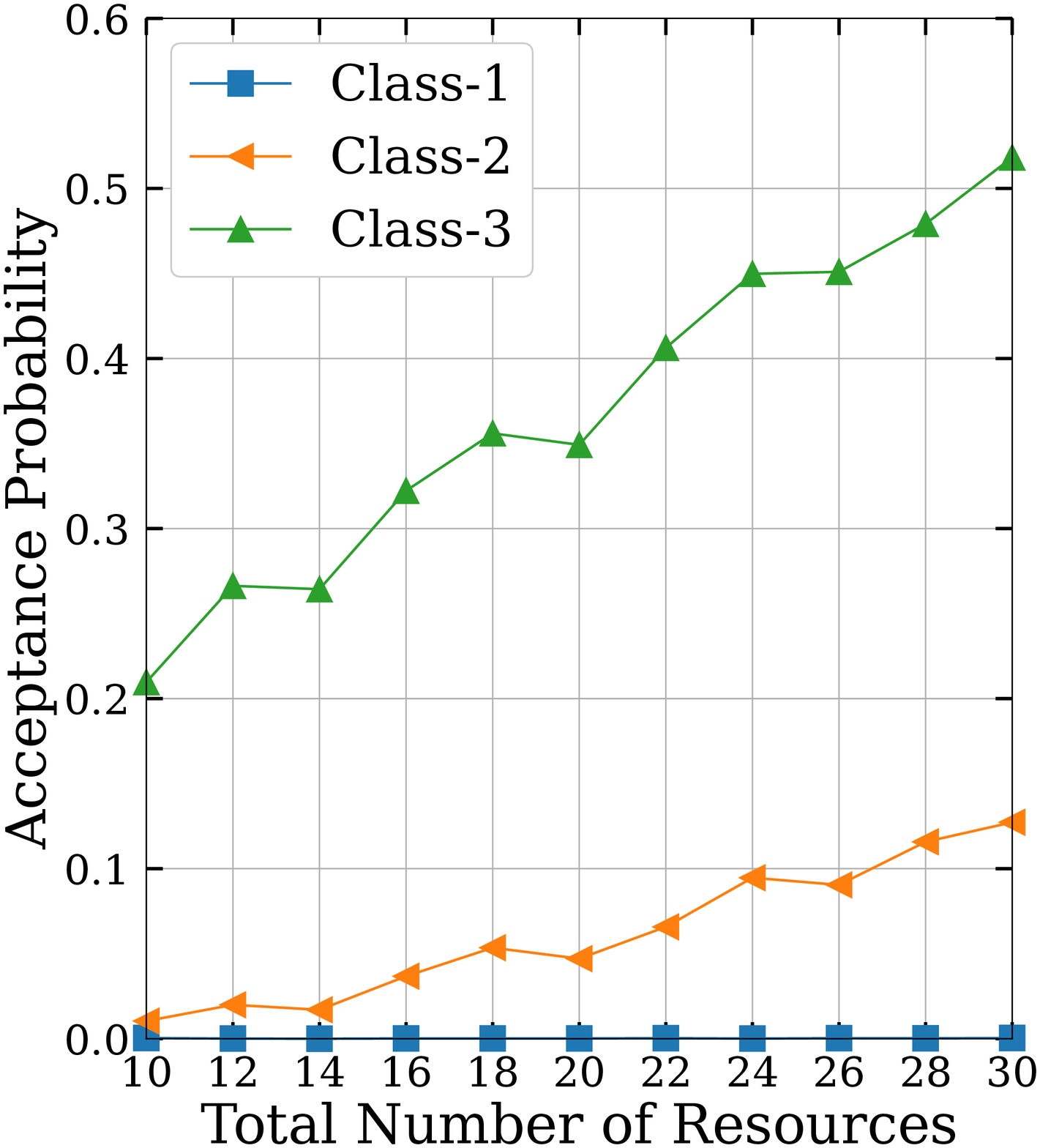} 
%		&\includegraphics[width=0.25\linewidth]{Figures/fig_vms_APbyClass_Greedy_MA}
		&\includegraphics[width=0.25\linewidth]{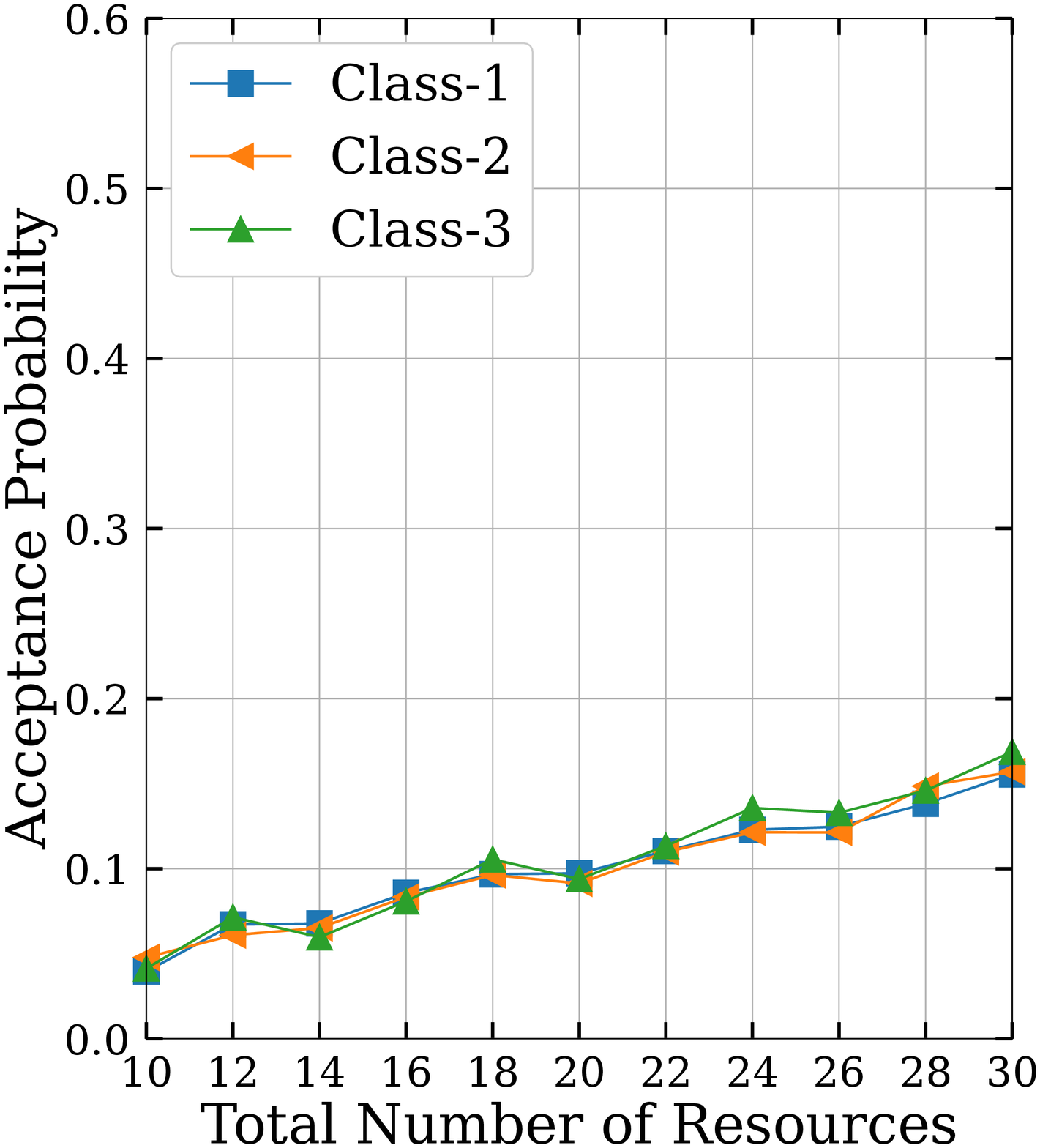}\\
		\text{(a) iMSAC+MA}&\text{(b) iMSAC}&\text{(c) Greedy}
	\end{array}$
	\caption{The acceptance probability per class when varying total number of resources.}
	\label{fig:vary_vms2}
	\vspace{-16pt}
\end{figure*}

The parameters of the iMSAC algorithm are set as follows.
	The value of $\epsilon$ is first set to one, and then it slowly decreases to $0.001$.
	We set the discount factor $\alpha$ to $0.9$.
	The hyperparameters of Q-network are set similar to those in~\cite{wang_dueling_2016}, e.g., learning rate is $10^{-3}$, and $C\!=\!10^4$. 	
	Recall that the proposed solution includes the self-learning algorithm iMSAC and the MetaSlice analysis.
	Without the MetaSlice analysis, MetaInstances cannot be created, and thus no function is shared among ongoing MetaSlices.
	Hence, to evaluate our proposed solution, namely iMSAC+MA, we use two baseline methods, i.e., iMSAC and Greedy policy policy (that always accepts requests if the system has sufficient resources).

\subsection{Simulation Results}	

\label{sec:results}
Fig.~\ref{fig:convergene} shows the convergence rate of iMSAC in two schemes, i.e., with and without the MetaSlice Analyzer.
	The average reward obtained by the Greedy policy is also shown as a baseline.
	In this experiment, the radio, storage, and computing resources are set to $480$ MHz, $480$ GB, and $480$ GFLOPS/s, respectively.	
	It can be observed that although the iMSAC+MA converges slower than the iMSAC does, the average reward of the policy obtained by iMSAC+MA (i.e., $0.264$) is $48\%$ greater than that of the iMSAC (i.e., $0.17$).
	In addition, the average rewards of iMSAC+MA and iMSAC are $230\%$ and $123\%$ greater than that of the Greedy, respectively.

Next, we investigate the robustness of our proposed solution by varying the radio, storage, and computing resources from  400 MHz, 400 GB, and 400 GFLOPS/s to 1200 MHz, 1200 GB, and 1200 GFLOPS/s, respectively. 
	In other words, the maximum number of functions simultaneously running on the system is varied from $10$ to $30$.
	The policies learned by the iMSAC and iMSAC+MA are acquired after \mbox{$3.75\!\times\!10^5$} iterations.
	In this experience, we use average reward and acceptance probability metrics because they clearly demonstrate the system performance in terms of provider's long-term average revenue and the users' QoS, i.e., service availability.
	Fig.~\ref{fig:vary_vms} demonstrates that the average rewards and acceptance probabilities obtained by all solutions increase when the system resource quantity increases.
	It is stemmed from the fact that with more resources, the system can host a greater number of MetaSlices, and thus it can accept more requests to get higher revenue than those of the system with a lower amount of resources.
	As the amount of system resources increases, iMSAC+MA always obtains the best results in terms of average reward and acceptance probability, which are up to $120\%$ and $178.9\%$ greater than those of the iMSAC, respectively, as shown in Fig.~\ref{fig:vary_vms}.
	Interestingly, even though the acceptance probability of the iMSAC is slightly lower than that of the Greedy, the iMSAC consistently achieves higher average rewards (e.g., up to $182\%$) than the Greedy does, as shown in Fig.~\ref{fig:vary_vms}(a).
	The reason can be observed in Fig.~\ref{fig:vary_vms2} where we look deeper at the acceptance probability per class.
	Specifically, in the iMSAC+MA and iMSAC, the acceptance probability of class-3, which has the highest immediate reward (i.e., $4$), is much higher than that of class-1, which has the lowest immediate reward, i.e., $1$.
	In contrast, in the Greedy, the acceptance probabilities of all classes are similar.
	The above results clearly demonstrate the effectiveness and robustness of our proposed solution.

%%%%%%%%%%%%%%%%%%%%%%%%%%%%%%%%%%%%%%%%%%%%%%%%%%%%%%%%%%%%%

\section{Conlcusion}
This paper has proposed the novel resource management framework to dynamically allocate appropriate resources from different tiers for effectively addressing the massive resource demands of Metaverse's applications, thereby maximizing the provider's long-term revenue.
To capture real time, dynamic and uncertainty of request arrival and MetaSlice departure processes, we have developed the sMDP-based framework and the intelligent algorithm that can gradually learn the optimal admission policy without requiring complete information about these processes.
The simulation results clearly show the effectiveness and robustness of our proposed solution.

%%%%%%%%%%%%%%%%%%%%%%%%%%%%%%%%%%%%%%%%%%%%%%
%\bibliographystyle{IEEEtran}
%\bibliography{refs_globecom}

\end{document}